\documentstyle{elsart}
\begin{document} 
\begin{frontmatter}
\everymath={\displaystyle}

\title{Phase transition and thermodynamics of a hot and dense system in
a scaled NJL model}

\author{J. Cugnon, M. Jaminon and B. Van den Bossche}

\address{Universit\'{e} de Li\`{e}ge, Institut de Physique B5, Sart
Tilman, B-4000 Li\`{e}ge 1, Belgium}

\begin{abstract} 
The chiral phase transition of a hot and dense system
of quarks is studied within a modified SU(3) NJL lagrangian that
implements the QCD scale anomaly.  The $u$- and $s$-quark condensates
can feel or not the same chiral restoration depending on the considered
region of the 3-dimension space
$\left({{T}_{c}\ ,\ {\mu }_{uc}\ ,\ {\mu }_{sc}}\right)$.  The
temperature behaviour of the pressure and of the energy and entropy
densities of the $u$- and $s$-quark system is investigated.  At high
temperature, the non-vanishing bare $s$-quark mass only modifies
slightly the usual behaviour associated with an ideal quark gas.
\end{abstract}


\end{frontmatter}
\pagebreak

\section{INTRODUCTION} 
\label{intro}
It is generally accepted that QCD generates two
phase transitions, a deconfinement one, corresponding to a transition
of a hadron gas to a quark-gluon plasma and a chiral one, corresponding
to the passage from a chiral symmetry breaking phase to a phase where
chiral symmetry is restored.  Lattice QCD calculations, based on the
estimate of the Polyakov loops, show that the two transitions coincide
in the pure gauge case.  Lattice calculations including massive quarks
\cite{CH92} (albeit not yet in a perfect shape) seem to indicate that
this situation persists \cite{DE87a,DE87b,GO87} after the introduction
of quarks.  However the order of the transition is changing, from first
order, for the pure gauge case, to second order with reasonable values
of the quark masses.  The case of the usually accepted values for the
bare $u$, $d$ and $s$ masses seems to be close to the boundary
separating first and second order transitions at zero quark chemical
potential.  Nothing is really known for non-zero chemical potential.\par

Whereas the deconfinement cannot be studied in existing satisfying
models, the features of spontaneous chiral symmetry breaking and of its
restoration can be studied in the $\sigma$-model or in the Nambu
Jona-Lasinio (NJL) model
\cite{NA61}.  The latter seems to be very successful in the description
of many features of hadronic physics in the non-perturbative domain of
QCD, despite of the absence of confinement and of renormalizability. 
It seems that the chiral aspects are much more important than the
confinement features, for static properties at least.  This model is
thus well suited to study the chiral transition of an infinite uniform
quark system, and the thermodynamics of the phases.  The pure NJL model
with SU(2) flavour has been used for this purpose by many authors
\cite{JA89,BE88,CH90,BE87,AS89,ZH94}.  Depending upon the unavoidable
choice for free parameters left in this kind of model, the transition
can be of first or second order \cite{JA94}, with a critical
temperature of the order of 150-200 MeV at zero chemical potential. 
Most of these approaches are restricted to the classical (mean field)
approximation.  Let us nonetheless notice the works of Zhuang et al.
\cite{ZH94} and of Blascke et al. \cite{BL95}, who calculated the
thermodynamic potential to the order
$\frac{1}{N_c}$, where
$N_c$ is the number of colors, which amounts roughly to include the
meson degrees of freedom.  The latter seem to play an important role at
low temperature only and does not affect significantly the position and
the properties (order, discontinuities,...) of the transition
\cite{ZH94}.  The chiral transition has been studied for the SU(3)
flavour case at zero chemical potential
\cite{HA87a,HA87b,HA94,JA95} and at non-zero light quark ($u$,$d$)
chemical potential
\cite{KL90}.\par 

Here, we want to make a systematic study of the chiral
phase transition features for the SU(3) flavour case for any value of
the light quark $\mu_u$ (taken equal to $\mu_d$) and of the strange
quark $\mu_s$ chemical potentials.  For this, we use an extension of
the NJL model, which implements the trace anomaly of QCD through the
introduction of a scalar dilaton field, and the axial anomaly.  The
former allows the description of the gluon condensate and the latter is
dictated by the low energy hadron phenomenology.  The model as well as
its predictions for the temperature and density dependences of the
meson masses are described in ref. \cite{JA95}.  We will evaluate the
thermodynamic potential in the mean field approximation within this
model and make a systematic survey of the transition in the $T$,
$\mu_u$ and $\mu_s$ spaces.  The main thermodynamic quantities will be
derived for each phase.  We use the assumption of a vanishing current
mass for the light quarks.  The non-zero current $s$-quark mass
constitutes the only flavour SU(3) symmetry breaking.\par 

The paper is organized as follows.  The NJL model used here is
schematically described in section~\ref{section2}, as well as the
associated action.  Section~\ref{section3} is devoted to the
localisation of the chiral phase transition in the $T$,
$\mu_u$ and
$\mu_s$ space for two choices of the free parameters of the model.  For
the first one, the transition is only of first order, whereas the
second one allows the possibility of a second order transition. 
Section~\ref{section4} deals with the pressure, energy and entropy
densities in the different phases.  The limiting values in the chiral
phase are particularly examined.  Section~\ref{section5} contains our
conclusion.

\section{THE A-SCALING NJL MODEL} 
\label{section2}
The model used in this paper is extensively described in Ref.
\cite{JA95} under the name ``A-scaling model".  We just recall some of
the tools useful to the understanding of the present paper.\par 

We start from the SU(3) effective Euclidean action :
\begin{eqnarray}
{I}_{eff}\left({\varphi,\chi,\mu,\beta
}\right)&=&-{Tr}_{\Lambda\chi}\ln\left({-i\partial\!\!\!/-W\!\!\!\!\!/\
+m+{\Gamma }_{a}{\varphi}_{a}}\right) \nonumber\\
 &+&\int{d}^{4}x{\frac{{a}^{2}{\chi}^{2}}{2}}{\varphi }_{a}{\varphi
}_{a}+\int{d}^{4}x{\cal {L}}_{\chi }+\int{d}^{4}x{\cal
 {L}}_{{U}_{A}(1)}
\label{eq1}
\end{eqnarray}
where the quark fields have been integrated out.  The meson fields
write :
\begin{equation} 
{\varphi}_{a}=\left({{\sigma }_{a},{\pi }_{a}}\right)\
\ \ ,\ \ \ {\Gamma }_{a}=\left({{\lambda }_{a},{i\gamma }_{5}{\lambda
}_{a}}\right)\ \ \ \ \ a=0,...,8
\label{eq2}
\end{equation} 
where $\lambda_{a}$ are the usual Gell-Mann matrices
$\left({{\lambda}_{0}=
\sqrt{{\frac{2}{3}}}1\!\!1}\right)$ and the gluon condensate is
represented by the scalar dilaton field
$\chi\propto{\left\langle{{G}_{\mu \nu }^{2}}\right\rangle}^{1/2}$. 
The lagrangian ${\cal {L}}_{\chi}$ given by
\begin{equation} 
{\cal {L}}_{\chi }={\frac{1}{2}}{\left({{\partial
}_{\mu }\chi }\right)}^{2}+{\frac{1}{16}}{b}^{2}\left({{\chi
}^{4}\ln{\frac{{\chi }^{4}}{{\chi }_{G}^{4}}}-\left({{\chi }^{4}-{\chi
}_{G}^{4}}\right)}\right)
\label{eq3}
\end{equation} 
takes account of the QCD trace anomaly via the term
${\chi}^{4}\ln{\chi}^{4}/{\chi}_{G}^{4}$ \cite{CA90,BR91}.  The
$U_{A}(1)$ axial anomaly is introduced via the mass term :
\begin{equation} 
{\cal {L}}_{{U}_{A}(1)}={\frac{1}{2}}{a}^{2}\xi {\chi
}^{ 2}{{\pi }_{0}}^{ 2} .
\label{eq4}
\end{equation} 
In Eq.~(\ref{eq1}), the trace has been regularized by an
ultra-violet cut-off which is
$\chi$-dependent in order to preserve the scale invariance of the quark
loop term. The chemical potentials are introduced via the vector field
${W}_{\lambda}=\left({-i\mu,\vec{0}}\right)$, where $\mu$ is a 
shorthand notation for
$\mu = diag\left(\mu_{\mu},\mu_{\mu},\mu_{s}\right)$, if one neglects
the isospin breaking $\left(\mu_{\mu}=\mu_{d}\right)$.  In the same way
the quantity $m$ represents the diagonal matrix
$m=diag\left(m_{u},m_{u},m_{s}\right)$ in flavor space.\par 

We work in the classical approximation so that the partition function
writes :
\begin{equation} 
Z\equiv {Z}_{c}=\exp\left[{-{I}_{eff}^{c}
\left({{M}_{u},{M}_{s},{\chi}_{s}}\right)}\right],
\label{eq5}
\end{equation} 
where $M_{u}\left(M_{s}\right)$ and $\chi_{s}$ denote
respectively the constituent $u$-quark ($s$-quark) mass and the gluon
condensate in the hot and dense system and are thus $\mu$- and
$\beta$-dependent ($\beta$ = 1/T).  In the vacuum, the corresponding
quantities are noted
$M_{u}^{0}$, $M_{s}^{0}$ and $\chi_{0}$.  These three quantities (in
the vacuum or in the system) minimize the thermodynamical potential or
equivalently the action (\ref{eq1}), yielding the value $I_{eff}^{c}$
entering eq.~(\ref{eq5}).  The relationship between
$M_{u}$,
$M_{s}$ and the stationary fields
$\sigma_{0}^{s}$, $\sigma_{8}^{s}$ is given in \cite{JA95}.  We write :
\begin{equation}
{I}_{eff}^{c}\left({{M}_{u},{M}_{s},{\chi}_{s}}\right)={I}_{\left({\mu,
\beta}\right)}^{c}\left({{M}_{u},{M}_{s},{\chi}_{s}}\right)+
{I}_{\left({0,\infty
}\right)}^{c}\left({{M}_{u},{M}_{s},{\chi}_{s}}\right)
\label{eq6}
\end{equation} 
where
\begin{eqnarray}
{I}_{\left({\mu,\beta}\right)}^{c}\left({{M}_{u},{M}_{s},{\chi}_{s}}
\right)=&-&Tr\ln{\left({-i\partial\!\!\!/-W\!\!\!\!\!/+m+{\Gamma}_{a}
{\varphi}_{a}}\right)}^{c}\nonumber\\
&+&Tr\ln{\left({-i\partial\!\!\!/+m+{\Gamma}_{a}
{\varphi}_{a}}\right)}^{c},
\label{eq7}
\end{eqnarray}
where the index $c$ indicates that the traces have to be evaluated
with the stationary values of the fields.  The contribution (\ref{eq7})
to the action is finite and writes :
\begin{eqnarray}
{I}_{\left(\mu,\beta\right)}^{c}
\left({M}_{u},{M}_{s},{\chi}_{s}\right)=&-&\beta\Omega
\frac{{2N}_{c}}{{2\pi}^{2}\beta}\sum\limits_{i=u,s}^{}{a}_{i}
\int_{0}^{\infty}{k}^{2}dk \nonumber\\
\times\left\{\ln\left(1+\exp\left[-\beta\left({E}_{i}+{\mu}_{i}
\right)\right]\right)\right.&+&\left.\ln\left(1+\exp\left[-\beta
\left({E}_{i}-{\mu}_{i}\right)\right]\right)\right\}
\label{eq8}
\end{eqnarray}
with ${E}_{i}=\sqrt{{k}^{2}+{{M}_{i}}^{2}}$, $a_{u}=2$, $a_{s}=1$ and
$\Omega$ the volume of the system.  The second term of the r.h.s.
of~(\ref{eq6}) corresponds to $\mu_u=\mu_s=0$,
$\beta=\frac{1}{T}=\infty$.  It writes :
\begin{eqnarray}
&{I}_{\left({0,\infty}\right)}^{c}\left({{M}_{u},{M}_{s},{\chi}_{s}}
\right)=-Tr_{{\Lambda\chi}_{s}}\ln{\left({-i\partial\!\!\!/+
m+{\Gamma}_{a}{\varphi}_{a}}\right)}^{c}&\nonumber\\
&+\int{d}^{4}x\left\{{{\frac{{a}^{2}{\chi}_{s}^{2}}{2}}\left[{{\left(
{{\sigma}_{0}^{s}}\right)}^{2}+{\left({{\sigma}_{8}^{s}}\right)}^{2}}
\right]+{\frac{1}{16}}{b}^{2}\left[{{\chi}_{s}^{4}\ln{\left(
{{\frac{{\chi}_{s}}{{\chi}_{G}}}}\right)}^{4}-\left({{\chi}_{s}^{4}-
{\chi}_{G}^{4}}\right)}\right]}\right\}&
\label{eq9}
\end{eqnarray}
which yields with a $d^{4}k$ cut-off :
\begin{eqnarray}
&{I}_{\left({0,\infty}\right)}^{c}\left({{M}_{u},{M}_{s},{\chi}_{s}}
\right)=-\beta\Omega{\frac{{2N}_{c}}{{32\pi}^{2}}}\sum
\limits_{i=u,s}^{}{a}_{i}&\nonumber\\ 
&\times\left\{{\left[{{\left({\Lambda
{\chi}_{s}}\right)}^{4}-{{M}_{i}}^{4}}\right]
\ln\left[{{\left({\Lambda{\chi}_{s}}\right)}^{2}+{{M}_{i}}^{2}}\right]
-\left[{{\frac{1}{2}}{\left({\Lambda{\chi}_{s}}\right)
}^{2}-{{M}_{i}}^{2}}\right]{\left({\Lambda
{\chi}_{s}}\right)}^{2}}+{{M}_{i}}^{4}\ln{{M}_{i}}^{2}
\right\}&\nonumber\\
&+\beta\Omega\left\{{{\frac{{a}^{2}{\chi}_{s}^{2}}{4}}
\sum\limits_{i=u,s}^{}{a}_{i}{{M}_{i}}^{2}{\left({1-{
\frac{{m}_{i}}{{M}_{i}}}}\right)}^{2}+
{\frac{{b}^{2}}{16}}\left[{{{\chi}_{s}}^{4}
\ln{\left({{\frac{{\chi}_{s}}
{{\chi}_{G}}}}\right)}^{4}-\left({{\chi}_{s}^{4}-{\chi}_{G}^{4}}
\right)}\right]}\right\}.&\nonumber\\
&&
\label{eq10}
\end{eqnarray}
Our model contains seven parameters : the bare quark masses
$\left(m_{u},m_{s}\right)$, the strengths $\left(a^{2}, b^{2}\right)$,
the vacuum gluon condensate $\chi_{0}$, the cut-off $\Lambda$ and the
parameter
$\xi$ of the axial anomaly.  Five of these seven parameters are fixed to
reproduce the vacuum masses of the pion $m_{\pi}$ = 139 MeV, of the
kaon $m_K$ = 496 MeV, of the $\eta'$ ${m}_{\eta'}$ = 958 MeV and of the
glueball
$m_{GL}$ = 1.3 GeV, together with the pion decay constant $f_{\pi}$ =
93 MeV.  The two other parameters remain free and they have been chosen
to be the vacuum constituent
$u$-quark mass $M_{u}^{0}$ (related to $a^2$) and the vacuum gluon
condensate
$\chi_0$.  We give results for two sets of parameters :
\begin{equation} 
{M}_{u}^{0}=600\ {\rm MeV}\ \ ,\ \ {\chi}_{0}=125\
{\rm MeV}
\label{eq11}
\end{equation}
\begin{equation} 
{M}_{u}^{0}=300\ {\rm MeV}\ \ ,\ \ {\chi}_{0}=80\ 
{\rm MeV}.
\label{eq12}
\end{equation} 
This choice is driven by the fact that at $\mu_{u}=\mu_{s}=0$, the phase
transition is of first order ($T_{c}=168$) MeV with the first set and
of second order ($T_{c}$=150 MeV) with the second set.  Note that in
the SU(2) case \cite{JA94},
$T_c$ was found to be 140 MeV with the second set, in agreement with
the lattice data \cite{KO91}.  We then investigate how finite density
can affect the order of the transition.  Since we are interested in the
determination of the parameters
$\left({{T}_{c},{\mu}_{uc},{\mu}_{sc}}\right)$ at the phase transition,
we choose to work in the limit $m_{u}=0$.

\section{CHIRAL PHASE TRANSITIONS}
\label{section3}
The quark condensates are related to the constituent masses in the
following way :
\begin{equation}
\left\langle{\overline{u}u}\right\rangle\equiv{\frac{1}{\beta\Omega}}
{\frac{{\partial I}_{eff}^{c}}{{\partial m}_{u}}}
\left({{M}_{u},{M}_{s},{\chi}_{s}}\right)=-{\frac{1}{2}}{a}^{2}
{\chi}_{s}^{2}{M}_{u}
\label{eq13}
\end{equation}
\begin{equation}
\left\langle{\overline{s}s}\right\rangle\equiv{\frac{1}{\beta\Omega}}
{\frac{{\partial I}_{eff}^{c}}{{\partial m}_{s}}}
\left({{M}_{u},{M}_{s},{\chi}_{s}}\right)=-{\frac{1}{2}} {a}^{2}{\chi
}_{s}^{2}\left({{M}_{s}-{m}_{s}}\right).
\label{eq14}
\end{equation} 
The coupling between these two condensates comes through
their coupling to the gluon condensate
$\chi_s$.  The value of $\chi_s$ is indeed coupled to those of $M_u$
and $M_s$ by the search of the minimum of the action~(\ref{eq6}) (see
Eq.~(\ref{eq1})).  This coupling can be better visualized by looking at
the ``three coupled gap equations" (see
\cite{JA95}) that amount to minimize the action.\par 

The chiral phase transitions happen when the quark condensates go to
zero.  There is a priori no reason why the two condensates should
vanish at the same time.  This is illustrated in Fig.~1 where the open
circles represent the phase transition $\mu_{sc}$ versus $\mu_{uc}$,
corresponding to a vanishing $u$-quark condensate at
$T_{c}=0$ (a), $T_{c}=100$ MeV (b), $T_{c}=161$ MeV (c), for the set of
parameters~(\ref{eq11}).  The transition for the $s$-quark is given by
the crosses.  At sufficiently small temperatures, one observes that the
chiral transition of the
$u$- and $s$-condensates are independent of one another (as in a pure
NJL model without 't Hooft determinant), except in a small region of
the space
$\left(\mu_{sc},\mu_{uc}\right)$ where the coupling via the gluon
condensate is sufficiently strong to force the two quark condensates to
go to zero at the same time.  When the temperature is high enough, only
this region of strong coupling is present.  Whatever the critical point
of the three-dimension space
$\left({{T}_{c},{\mu }_{uc},{\mu }_{sc}}\right)$, the transition is of
first order.  This is illustrated in Fig.~2 in the case of strong
coupling (between the condensates) (a) (${T}_{c}\approx 120$ MeV,
${\mu}_{uc}={\mu}_{sc}=250$ MeV) and of weak coupling (b) (${T}_{\mu
c}=140$ MeV, ${T}_{sc}=160$ MeV,
${\mu}_{uc}=200$ MeV, ${\mu}_{sc}=0$ MeV).  Within the present model, a
discontinuity (transition) of a quark condensate is always accompanied
by a discontinuity of the gluon condensate, in contradistinction with
the suggestion of Ref. \cite{BR93}.  Note that
${\chi}\ne 0$ above the transition.  Fig.~3 summarizes the previous
results in the form of an artistic view.  The space is divided into
four regions : the inner one corresponds to quarks in their condense
phase ; the outer one can be seen as the quark-gluon plasma, the two
remaining regions (extending along the axes to infinity) correspond to
mixed phases with free $u$ ($s$)-quarks together with
$s$ ($u$)-quarks in the hadronic phase if one assumes that the chiral
transition corresponds to deconfinement.  A remarkable prediction of
the present model has to be singled out : if non-strange matter at
sufficiently large up chemical potential is heated, the matter will
transform first to a mixture of free light quarks surrounding strange
hadrons (with total vanishing strangeness) before being transformed
into a quark gas.\par 

The densities
$\rho_u$ and
$\rho_s$ are related to the chemical potentials and temperature in the
following way
\begin{equation}
{\rho}_{i}=-{\frac{1}{\beta\Omega}}\left({{\frac{{\partial I}_{eff}^{c}}
{{\partial\mu}_{i}}}}\right)=-{\frac{{N}_{c}{a}_{i}}{{\pi}^{2}}}
\int_{0}^{\infty}{k}^{2}\left({{n}_{+i}-{n}_{-i}}\right)dk\ \ \ i=u,s
\label{eq15}
\end{equation} 
where
\begin{equation} 
{n}_{\pm
i}=\frac{1}{1+\exp\left\{\beta\left(\sqrt{k^2+M_i^2}\pm\mu_i\right)
\right\}}\ \ \ i=u,s.
\label{eq16}
\end{equation} 
In Eq.~(\ref{eq15}), $\rho_u$ takes into account the
$d$-quark contribution via
$a_{u}=2$. When the transition is of first order, there is a jump in
the density at the transition.  Just before the transition, the
$\rho_i$'s are given by~(\ref{eq15}) and~(\ref{eq16}), with
${M}_{i}^{2}\ne{m}_{i}^{2}$ (${\rho}_{i\ bef}$), while just after the
transition, the densities (${\rho}_{i\ aft}$) increase up to values
corresponding to $M_{i}=m_{i}$.  A drawback of the present model is
that at $T_{c}=0$, ${\rho}_{i\ bef}=0$, so that the system can not
exist in the hadronic phase.  As $T_c$ departs from zero, $\rho_{i\
bef}$ remains small.  As an example, for
${\mu}_{uc}={\mu}_{sc}=250$ MeV, one has ${\frac{{\rho}_{u\
bef}}{{\rho}_{0}}} =0.362$, ${\frac{{\rho}_{s\
bef}}{{\rho}_{0}}}=0.050$ and
${\frac{{\rho}_{u\ aft}}{{\rho}_{0}}}=2.646$, ${\frac{{\rho}_{s\ aft}}
{{\rho}_{0}}}=0.978$ where ${\rho}_{0}={0.51\ fm}^{-3}$.  Within the
present model, it is practically impossible to eliminate this drawback,
as the only way would correspond to having small constituent masses in
the hadronic phase while keeping large gluon condensate.  Such a choice
of parameters is not suitable \cite{JA94} since it would yield too high
$T_c$.  A hope lies however in the inclusion of vector mesons which are
expected to shift the critical densities up to acceptable values
without modifying the qualitative features of the phase transition
\cite{JA93}, as the latter are basically determined by the chiral
mesons and their coupling to the condensates.\par
 
The previous results are not qualitatively changed drastically when
shifting to the other choice of the parameters (Eq.~(\ref{eq12})),
except for the order of the transition which can become of second order
at sufficiently low chemical potentials.  This is illustrated in
Fig.~4a.  At
$\mu_u=\mu_s=0$, the $u$-quark condensate vanishes at
${T}_{uc}\approx150$ MeV (140 MeV in SU(2)) while chiral transition
happens at ${T}_{sc}\approx  200$ MeV for the $s$-quark.  At higher
chemical potentials, the transition becomes of first order but the
chiral transitions for the $u$- and $s$-quarks remain decoupled (see
Fig.~4b).  Note, however, that a change of the gluon condensate is
always associated with the transition of any quark condensate.  Figs.~5
and~6 are the analogs of Figs.~1 and~3, respectively for the
set~(\ref{eq12}).  Figs.~1 and~5 are qualitatively similar except at
high temperature where the two quark condensates are coupled whatever
the couple (${\mu}_{uc}$, ${\mu}_{sc}$) for the set~(\ref{eq11}) while
the $u$-quark can only exist in the ``deconfined" phase with a chiral
 transition at ${\mu}_{sc}\approx 220$ MeV whatever ${\mu}_{uc}$ for the
strange quark, in the case of  set~(\ref{eq12}).  The results presented
in Fig.~5 can be visualized in the 3-dimension space by use of the
artistic view of Fig.~6.

\section{THERMODYNAMICS}
\label{section4}
We now turn to the study of the thermodynamics of a hot and/or dense
system.  We will mainly focus on the behaviour with respect to the
temperature of the pressure, the energy density and the entropy
density.  These three quantities are respectively given in the
classical approximation by :
\begin{eqnarray} 
P&=&{\frac{1}{\beta}}{\frac{\partial\ln\left({{Z}_{c}/
{Z}_{c}^{0}}\right)}{\partial\Omega}}=-{\frac{1}{\beta}}
\left[{{\frac{{\partial I}_{eff}^{c}\left({{M}_{u},{M}_{s},{\chi}_{s}}
\right)}{\partial\Omega}}-{\frac{{\partial
I}_{\left(0,\infty\right)}^{c}
\left({{M}_{u}^{0},{M}_{s}^{0},{\chi}_{0}}\right)}{\partial\Omega}}}
\right]\nonumber\\
&=&-{\frac{1}{\beta\Omega}}\left[{{I}_{\left({\mu,\beta}\right)}^{c}
\left({{M}_{u},{M}_{s},{\chi}_{s}}\right)+{I}_{\left({0,\infty}
\right)}^{c}\left({{M}_{u},{M}_{s},{\chi}_{s}}\right)-{I}_{\left({0,
\infty}\right)}^{c}\left({{M}_{u}^{0},{M}_{s}^{0},{\chi}_{0}}\right)}
\right]\nonumber\\ 
&&
\label{eq17}
\end{eqnarray}
\begin{eqnarray}
\epsilon\equiv{\frac{E}{\Omega}}&=&{-\frac{1}{\Omega }}
\left({{\frac{\partial}{\partial\beta}}-{\frac{1}{\beta}}\sum
\limits_{i=u,s}^{}{\mu}_{i}{\frac{\partial}{{\partial\mu}_{ i}}}}
\right)\ln\left({{Z}_{c}/{Z}_{c}^{0}}\right)\nonumber\\
&=&{\frac{1}{\Omega}}\left({{\frac{\partial}{\partial\beta}}-
{\frac{1}{\beta}}\sum\limits_{i=u,s}^{}{\mu}_{i}{\frac{\partial}
{{\partial\mu}_{i}}}}\right)\left[{{I}_{eff}^{c}
\left({{M}_{u},{M}_{s},{\chi}_{s}}\right)-{I}_{
\left(0,\infty\right)}^{c}
\left({{M}_{u}^{0},{M}_{s}^{0},{\chi}_{0}}\right)}\right]\nonumber\\
&=&{\frac{1}{\Omega}}\left({{\frac{\partial}{\partial\beta}}-
{\frac{1}{\beta}}\sum\limits_{i=u,s}^{}{\mu}_{i}
{\frac{\partial}{{\partial\mu}_{i}}}}\right){I}_{\left({\mu,\beta}
\right)}^{c}\left({{M}_{u},{M}_{s},{\chi}_{s}}\right)\nonumber\\
&+&{\frac{1}{\beta\Omega}}\left[{{I}_{\left({0,\infty}\right)}^{c}
\left({{M}_{u},{M}_{s},{\chi}_{s}}\right)-{I}_{
\left({0,\infty}\right)}^{c}
\left({{M}_{u}^{0},{M}_{s}^{0},{\chi}_{0}}\right)}\right]
\label{eq18}
\end{eqnarray}
\begin{eqnarray}
s\equiv\frac{S}{\Omega}&=&\beta\left({P+{\frac{E}{\Omega}}-\sum
\limits_{i=u,s}^{}{\mu}_{i}{\rho}_{i}}\right)\nonumber\\
&=&-{\frac{1}{\Omega}}
\left({1-\beta{\frac{\partial}{\partial\beta}}}\right){I}_{\left({\mu,
\beta}\right)}^{c}\left({{M}_{u},{M}_{s},{\chi}_{s}}\right),
\label{eq19}
\end{eqnarray} 
where we have substracted the vacuum contribution
$Z_{c}^{0}$ from the partition function in order to ensure vanishing
pressure, energy and entropy in the vacuum.\par 

Following the same prescription as in \cite{KO93}, we can write
\begin{equation} 
P={P}_{ideal\ gas}-B\left({\beta,{\mu}_{u},{\mu}_{s}}\right)
\label{eq20}
\end{equation}
\begin{equation}
\epsilon={\varepsilon}_{ideal\
gas}+B\left({\beta,{\mu}_{u},{\mu}_{s}}\right)
\label{eq21}
\end{equation}
\begin{equation} 
Ts={P}_{ideal\ gas}+{\varepsilon}_{ideal\ gas}-\sum
\limits_{i=u,s}^{}{\mu}_{i}{\rho}_{i}
\label{eq22}
\end{equation} 
with
\begin{equation} 
{P}_{ideal\ gas}=-{\frac{{I}_{\left({\mu,\beta}\right)}^{c}
\left({{M}_{u},{M}_{s},{\chi}_{s}}\right)}{\beta\Omega}}=
{\frac{{N}_{c}}{{3\pi}^{2}}}\sum\limits_{i=u,s}^{}{a}_{i}
\int_{0}^{\infty}{\frac{{k}^{4}}{{E}_{i}}}\left({{n}_{+i}+{n}_{-i}}
\right)dk
\label{eq23}
\end{equation}
\begin{eqnarray} 
{\varepsilon}_{ideal\ gas}&=&{\frac{1}{\Omega}}
\left({{\frac{\partial}{\partial\beta}}-{\frac{1}{\beta}}\sum
\limits_{i=u,s}^{}{\mu}_{i}{\frac{\partial}{{\partial
\mu}_{i}}}}\right){I}_{\left({\mu,\beta}\right)}^{c}
\left({{M}_{u},{M}_{s},{\chi}_{s}}\right)\nonumber\\
&=&{\frac{{N}_{c}}{{\pi}^{2}}}\sum\limits_{i=u,s}^{}{a}_{i}
\int_{0}^{\infty}{k}^{2}{E}_{i}\left({{n}_{+i}+{n}_{-i}}\right)dk
\label{eq24}
\end{eqnarray}
and $B$ denotes the (temperature and density-dependent) bag pressure
\begin{equation}
B\left({\beta,{\mu}_{u},{\mu}_{s}}\right)={\frac{1}{\beta\Omega}}
\left\{{{I}_{\left({0,\infty}\right)}^{c}
\left({{M}_{u},{M}_{s},{\chi}_{s}}
\right)-{I}_{\left({0,\infty}\right)}^{c}
\left({{M}_{u}^{0},{M}_{s}^{0},{\chi}_{0}}\right)}\right\}.
\label{eq25}
\end{equation} 
The definition~(\ref{eq25}) differs from the one used in
\cite{JA95,CA90,BR91} where the bag pressure is defined as:
\begin{eqnarray} 
B&=&{\frac{1}{\beta\Omega}}
\left\{{{I}_{\left({0,\infty}\right)}^{c}\left({0},{0},{0}\right)-
{I}_{\left({0,\infty}\right)}^{c}
\left({{M}_{u}^{0},{M}_{s}^{0},{\chi}_{0}}
\right)}\right\}\nonumber\\ 
&=&{\frac{1}{16}}{m}_{GL}^{2}{\chi}_{0}^{2}.
\label{eq26}
\end{eqnarray}
The temperature behaviour of $P$ is exhibited in Fig.~7a, for the two
sets of parameters~(\ref{eq11}) and~(\ref{eq12}) in the case
$\mu_{u}=\mu_{s}=0$.  Due to the non-vanishing value of the strange
quark bare mass
$m_s$, the temperature behaviour above $T_c$ is not exactly the usual
law in
$T^4$. One can write (see Appendix A)
\begin{eqnarray} 
{P}_{ideal\ gas}&\approx&{\frac{7}{60}}{N}_{c}{\pi}^{2}{T}^{4}
-{\frac{{N}_{c}}{12}}{m}_{s}^{2}{T}^{2}-
{\frac{{N}_{c}}{{8\pi}^{2}}}{m}_{s}^{4}\ln{\frac{{m}_{s}}{\pi
T}}\nonumber\\
&+&{\frac{{N}_{c}}{{16\pi}^{2}}}\left({{\frac{3}{2}}-2\gamma}\right)
{m}_{s}^{4}+\ ...,
\label{eq27}
\end{eqnarray}
where $\gamma$ is the Euler constant.  At the same time, $B$ becomes
$T$-independent
\begin{equation}
B={\frac{1}{\beta\Omega}}\left\{{{I}_{\left({0,\infty}\right)}^{c}
\left({0,{m}_{s},\chi_s}\right)-{I}_{\left({0,\infty}\right)}^{c}
\left({{M}_{u}^{0},{M}_{s}^{0},{\chi}_{0}}\right)}\right\},
\label{eq28}
\end{equation} 
where $\chi_s$ is the stationary value for $T\ >\
{T}_{c}$.  Fig.~7a shows an excellent $T^4$ linear fit of
$P$ above chiral restoration :
\begin{equation} 
P\approx 3.40\ {T}^{4}-2.097\ {10}^{-3},
\label{eq29}
\end{equation} 
for the set of parameters~(\ref{eq11}) and
\begin{equation} 
P\approx 3.40\ {T}^{4}-1.195\ {10}^{-3},
\label{eq30}
\end{equation} 
for the set~(\ref{eq12}). The coefficient of $T^4$ is
not exactly equal to the coefficient of the same power in~(\ref{eq27}),
which is equal to 3.454.  In fact, due to the limited range of $T^4$
investigated in Fig.~7, the remaining terms in Eq.~(\ref{eq27}) change
the apparent slope and the constant term of the curve. When a fit with
the explicitly given terms of Eq.~(\ref{eq27}) plus an adjusted
constant $C$ is performed, the latter turns out to be ${C}^{1/4}\approx
207$ MeV for set~(\ref{eq11}) and ${C}^{1/4}\approx 179$ MeV for
set~(\ref{eq12}).  These values are close to ${B}^{1/4}\approx 209$ MeV
(set~(\ref{eq11})) and ${B}^{1/4}\approx 183$ MeV (set~(\ref{eq12}))
obtained from a direct calculation using Eq.~(\ref{eq28}).  This
underlines the consistency and the precision of our numerical results as
well as the rapid convergence of expansion~(\ref{eq27}), although the
expansion parameter
$m_{s}/T$ is not very small.  Moreover, note that the numerical values
of
$B^{1/4}$ given by~(\ref{eq28}) or extracted from the fits are not so
far from the values of $B^{1/4}$ given by~(\ref{eq26}) ($B^{1/4}=201$
MeV (set~\ref{eq11}), 161 MeV (set~\ref{eq12})).\par 

Below the phase transition, the temperature behaviour depends on the
order of the transition.  With the set~(\ref{eq11}) (first order),
$\beta M_i$ ($i = u,s$) remain large, so that the $T$ behaviour is
governed by the law
\begin{equation} 
{P}_{ideal\ gas}\approx{\frac{{4N}_{c}{\beta}^{-5/2}}{{\left({2\pi}
\right)}^{3/2}}}\sum\limits_{i=u,s}^{}{a}_{i}{M}_{i}^{3/2}{e}^{- {\beta
M}_{i}}
\label{eq31}
\end{equation} 
It should be noted that the masses $M_i$ are decreasing
with temperature, which makes the variation of the pressure more rapid
that the $T^{5/2}$ factor in Eq.~(\ref{eq31}).  The masses $M_i$ being
larger with set~(\ref{eq11}) than with set~(\ref{eq12}), the pressure
increases more slowly in the first case and the change of slope at the
transition is more pronounced.\par 

The $T^4$ law for $P_{ideal\ gas}$ does not contain the explicit gluon
degrees of freedom which would yield :
\begin{equation} 
{P}_{ideal\
gas}\approx{\frac{{\pi}^{2}}{90}}\left\{{{2N}_{g}+{\frac{7}{8}}{N}_{c}
{N}_{f}4}\right\}{T}^{4}\approx 5.2\ {T}^{4}.
\label{eq32}
\end{equation} 
A first step in the inclusion of the gluon
thermodynamics consists in going beyond the classical approximation by
adding one loop corrections. One then has to add
\begin{eqnarray}
{P}_{one\ loop}&=&-\sum\limits_{j}^{}\int{\frac{{d}^{3}k}
{{\left({2\pi}\right)}^{3}}}\ln\left[{1-\exp\left({-\beta\sqrt{{k}^{2}+
{{m}_{j}}^{2}}}\right)}\right]\nonumber\\
&=&-\sum\limits_{j}^{}{\frac{1}{3}}\int{\frac{dk}{{2\pi}^{2}}}
{\frac{{k}^{4}}{\sqrt{{k}^{2}+{{m}_{j}}^{2}}}}{\frac{1}
{\left({1-\exp\beta\sqrt{{k}^{2}+{{m}_{j}}^{2}}}\right)}}
\label{eq33}
\end{eqnarray}
where the sum runs over the 19 mesons : (9+1) scalar and 9
pseudo-scalar mesons.  When chiral symmetry is restored, all the meson
masses ($m_j$) (except for the dilaton) increase to infinity for
increasing temperature so that the product
$\beta m_j$ remains large, and :
\begin{eqnarray} 
{P}_{one\ loop}(meson)\ &\approx&\sum\limits_{j}^{}
{\frac{1}{3}}\int{\frac{dk}{{2\pi}^{2}}}{\frac{{k}^{4}}{\sqrt{{k}^{2}+
{{m}_{j}}^{2}}}}{e}^{-\beta\sqrt{{k}^{2}+{{m}_{j}}^{2}}}\nonumber\\
&\approx&{\frac{{\beta}^{-5/2}}{{\left({2\pi}\right)}^{3/2}}}\sum
\limits_{j}^{}{{m}_{j}}^{3/2}{e}^{-\beta{m}_{j}}
\label{eq34}
\end{eqnarray}
in the non-relativistic limit.  The exponential factor makes the one
loop correction terms due to the 18 mesons vanishingly small compared
to the
$T^4$ term.  The mass $m_D$ of the 19th meson, the dilaton, remains
finite above the phase transition so that $\beta m_{D} \rightarrow 0$,
and
\begin{equation} 
{P}_{one\ loop}(dilaton)\approx{\frac{{\pi}^{2}}{90}}{T}^{4}+\ ...
\label{eq35}
\end{equation} 
where only one degree of freedom for the gluon is
included.\par 

A second step in the description of the gluon thermodynamics would
consist in including a temperature-dependent gluon potential
$V_{\chi}(T)$ as stipulated in
\cite{CA90,KU92}, which could yield the right asymptotic
behaviour~(\ref{eq32}).\par
 
Fig.~7b is the equivalent of Fig.~7a for $\mu_u=\mu_s=250$ MeV.  These
values of the chemical potentials correspond to a strong coupling
between
$\left\langle{\overline{u}u}\right\rangle$ and
$\left\langle{\overline{s}s}\right\rangle$ with the set~(\ref{eq11})
while they correspond to a weak coupling with the set~(\ref{eq12}). 
Even at high temperature, it is no more possible to perform a $T^4$
linear fit of the results.  Instead, the chemical potentials introduce
a $T^2$ dependence in the pressure.  In the chiral limit
(${m}_{u}={m}_{s}=0$), one would have
\begin{equation}
P={\frac{7}{60}}{N}_{c}{\pi}^{2}{T}^{4}+{\frac{{N}_{c}}{2}}{\mu}^{2}
{T}^{2}+{\frac{{N}_{c}{\mu }^{4}}{{4\pi}^{2}}}-B
\label{eq36}
\end{equation} 
with $\mu\equiv{\mu}_{u}={\mu}_{s}$.  Due to
${m}_{s}\ne0$, Eq.~(\ref{eq36}) has
 to be modified.  However, the prescription used in the case
$\mu=0$ (see Appendix~A) does not work here if ${\mu}_{i}>{m}_{i}$. 
From our results, one can see that the $T^4$ term is not modified by
$m_s$, as expected, and the $T^2$ term is slightly decreased.\par 

The temperature behaviour of the energy
density~(\ref{eq21}),~(\ref{eq24}) is exhibited in Fig.~8a, for
${\mu}_{u}={\mu}_{s}=0$.  At high $T$, one can write
\begin{eqnarray} 
{\varepsilon}_{ideal\ gas}\approx {\frac{7}{20}}
{N}_{c}{\pi}^{2}{T}^{4}&-&{\frac{{N}_{c}}{12}}{m}_{s}^{2}{T}^{2}+
{\frac{{N}_{c}}{{8\pi}^{2}}}{m}_{s}^{4}\ln{\frac{{m}_{s}}{\pi
T}}\nonumber\\
&+&{\frac{{N}_{c}}{{16\pi}^{2}}}\left({2\gamma+{\frac{1}{2}}}\right)
{m}_{s}^{4}\ ...
\label{eq37}
\end{eqnarray}
while the obtained results seem to show a $T^4$ linear behaviour :
\begin{equation}
\varepsilon\approx 10.269\ {T}^{4}+1.620\ {10}^{-3},
\label{eq38}
\end{equation}
\begin{equation}
\varepsilon\approx 10.271\ {T}^{4}+0.877\ {10}^{-3}
\label{eq39}
\end{equation} 
for sets~(\ref{eq11}) and~(\ref{eq12}), respectively. 
When a fit is performed with the terms written in Eq.~(\ref{eq37}) plus
an adjustable constant $C$, one obtains
${C}^{1/4}\approx 208$ MeV for set~(\ref{eq11}) and
${C}^{1/4}\approx 182$ MeV for set~(\ref{eq12}), in agreement with the
results obtained from the $T$ behaviour of the pressure.\par 

The first order phase transition and the strong coupling between
$\left\langle{\overline{u}u}\right\rangle$ and
$\left\langle{\overline{s}s}\right\rangle$ (in the case of the
set~(\ref{eq11})) is well visualized by the sudden change of the energy
value at $T_c$.  For set~(\ref{eq12}), the weak coupling between the
quark condensates and the presence of a second order phase transition
for the
$u$-quark (the $s$-quark still feeling a first order transition)  is
reflected by a slope breaking in $\varepsilon (T)$ at $T_{uc}$ and a
small jump at $T_{sc}$.\par 

Fig.~8b is equivalent of Fig.~8a in the case of a dense strange matter
${\mu}_{u}={\mu}_{s}=250$ MeV.  Here again, the chemical potentials
introduce a $T^2$ dependence which writes with $\mu\equiv{\mu }_{s}=
{\mu}_{u}$ :
\begin{equation}
\varepsilon={\frac{7}{20}}{N}_{c}{\pi}^{2}{T}^{4}+{\frac{3}{2}}
{N}_{c}{\mu}^{2}{T}^{2}+{\frac{3}{4}}{N}_{c}{
\frac{{\mu}^{4}}{{\pi}^{2}}}+B
\label{eq40}
\end{equation} 
in the chiral limit.  From our results, one sees that
${m}_{s}\ne 0$ only slightly affects the second term of~(\ref{eq40}). 
It also introduces an additional constant term.\par 

For a pure ideal gas, the entropy density behaves as $T^3$ at
$\mu=0$.  When the non-vanishing $s$-quark mass is taken into account,
one can use Eqs.~(\ref{eq22}),~(\ref{eq27}) and~(\ref{eq37}) to get the
three first terms of the expansion of $Ts$ in power of $T^2$.  One then
has for $s$ :
\begin{equation} 
s\approx{\frac{7}{15}}{N}_{c}{\pi}^{2}{T}^{3}-
{\frac{{N}_{c}}{6}}{m}_{s}^{2}T+{\frac{{N}_{c}}{{8\pi}^{2}}}
{m}_{s}^{4}{T}^{-1}+...,
\label{eq41}
\end{equation} 
where the dots represent more negative powers of $T$.
However, the results of Fig.~9a show that one can
approximate~(\ref{eq41}) by a
$T^3$ linear expression :
\begin{equation} 
s\approx 13.868\ {T}^{3}-0.136\ {10}^{-1},
\label{eq42}
\end{equation} 
\begin{equation} 
s\approx 13.839\ {T}^{3}-0.912\ {10}^{-2},
\label{eq43}
\end{equation} 
for sets~(\ref{eq11}) and~(\ref{eq12}) respectively. 
The coefficients of $T^3$ are not exactly equal to what would be
obtained from combining Eqs.~(\ref{eq29}),~(\ref{eq30}),~(\ref{eq38})
and~(\ref{eq39}), although the numerical results are perfectly
verifying Eq.~(\ref{eq19}).  Similarly, the constants in
Eqs.~(\ref{eq42}) and~(\ref{eq43}) are departing numerically from the
second and third terms of Eq.~(\ref{eq41}).  This reflects the limited
meaning of simple fits like Eqs.~(\ref{eq29}),~(\ref{eq38})
and~(\ref{eq42}) in a restricted range of
$T^4$.  This remark would apply to any calculation.  It would then be
dangerous to try to extract a bag constant from for instance $P$ versus
$T^4$ curve by a simple fit.\par 

Fig.~9b shows the temperature behaviour of $s$ in the case of
non-vanishing chemical potentials.  Using~(\ref{eq22}),~(\ref{eq36})
and~(\ref{eq40}) together with the density after chiral restoration
\begin{equation}
\rho\mu={\frac{{N}_{c}{\mu}^{4}}{{\pi}^{2}}}+{N}_{c}{\mu}^{2}{T}^{2}
\label{eq44}
\end{equation} 
one gets for an ideal gas :
\begin{equation}
Ts={\frac{7}{15}}{N}_{c}{\pi}^{2}{T}^{4}+{N}_{c}{\mu}^{2}{T}^{2}
\label{eq45}
\end{equation} 
expression which should be modified in order to take
${m}_{s}\ne 0$ into account.  We observed that this non-vanishing $m_s$
only slightly affects the value of the factor multiplying $T^2$.  It
should be noticed that above the transition, the entropy shows that all
the available degrees of freedom are excited.  However, as noted in
Refs. \cite{ZH94,LU92}, the massless Stefan-Boltzmann limit of QCD is
not reached, because not all the gluon degrees of freedom are manifest
in this model, and because the mass of the strange quark is not yet
negligible.

\section{CONCLUSION} 
\label{section5}
We have investigated the properties of the chiral
transition in an infinite system of quarks with the help of a SU(3)
flavour NJL model, built to account for the axial and scale anomalies. 
There is some freedom in chosing the parameters of the model.  We
illustrated our calculations for two choices, corresponding to large
and moderate constituent quark masses.  With the first choice, the
transition is always of first order.  With the second choice, the
transition can be of second order for small $u$ and $s$ chemical
potentials.  This is to be contrasted with many other NJL models (see
e.g. Refs.
\cite{HA94,KL90}), where the precise form of the Lagrangian, the way of
regularization and the choice of the parameters hardly allow for a
first order transition.  Let us mention that the chiral transition for
the flavour SU(3) case is also studied in ref. \cite{ME94} in a
$\sigma$-model and in ref.
\cite{HU95} in a NJL model.  These works however concentrate on the zero
chemical potential case and on the modification of the meson masses. 
They include $\frac{1}{N_f}$ \cite{ME94} and $\frac{1}{N_c}$
\cite{HU95} terms.\par 

An important feature of our results lies in the possibility of having
coupled, i.e. simultaneous, chiral transitions for light and strange
quarks.  These transitions tend to become uncoupled if either the
chemical potential $\mu_u$ or the chemical potential $\mu_s$ is large
enough.  We remind that, in our model, the
$u$ and $s$ condensates are only coupled indirectly through their
coupling to the gluon condensate.  As a consequence the latter is
always modified in a quark ($u$ or $s$) chiral transition.\par 

The scaled NJL model studied here allows the possibility of having first
order transitions, with important jumps of the thermodynamical
quantities.  This is an interesting feature in view of the search of
the quark-gluon plasma in heavy ion collisions.  However, as we
explained in section~\ref{section3}, the model should be made more
realistic through, e.g. the introduction of vector mesons.  It should
be checked how the discontinuities will be modified by this
introduction.\par
 
We also investigated the equation of state for the chiral phase.  The
latter is very close to the relativistic free gas one.  Departures from
the massless Stefan-Boltzmann limit comes from the non-vanishing mass
of the $s$ quark in the chiral phase and from the ``bag" constant,
which can be identified as due to the chiral symmetry breaking of the
vacuum (regardless of the non-zero current mass of the strange quark). 
Next steps in our investigations would involve the inclusion of the
mesons degrees of freedom, which can be viewed as the quantum
fluctuations of the quark fields \cite{JA95}, the study of their
influence on the equation of state, especially in chiral asymmetric
phase and the possibility of including all the gluon degrees of freedom.

\begin{ack} The work of B. Van den Bossche was supported by the
``Institut Interuniversitaire des Sciences Nucl\'eaires", Belgium.
\end{ack}

\section*{Appendix}

\begin{appendix}
\section{High temperature expansions for a hot ideal gas} We give
explicitly the expansion of thermodynamical quantities at zero chemical
potential and high temperature.
\begin{enumerate}
\item[a)] Pressure\\
\end{enumerate} 
At high $T$, the pressure of an ideal gas of massless
$u$ quarks and massive $s$ quarks is given by
\begin{equation} 
{P}_{ideal\ gas}={P}_{ideal\ gas}\left({u}\right)+{P}_{ideal\ gas}
\left({s}\right)
\label{eqA1}
\end{equation} 
or
\begin{equation} 
{P}_{ideal\
gas}=2{\frac{{N}_{c}}{{3\pi}^{2}}}\left\{{2\int_{0}^{\infty}
{\frac{{k}^{3}}{1+{e}^{\beta k}}}dk+\int_{0}^{\infty}
{\frac{{k}^{4}}{{E}_{s}}}{\frac{1}{1+{e}^{\beta{E}_{s}}}}dk}\right\}
\label{eqA2}
\end{equation}
with ${E}_{s}=\sqrt{{k}^{2}+{m}_{s}^{2}}$, $m_s$ being the bare
$s$-quark mass.  The first term of the r.h.s. of~(\ref{eqA2}) has the
usual
$T^4$ behaviour
\begin{equation} 
{P}_{ideal\ gas}(u)={\frac{7}{90}}{N}_{c}{\pi}^{2}{T}^{4}.
\label{eqA3}
\end{equation} 
The second term can be estimated in the following way :
\begin{equation} 
{P}_{ideal\ gas}(s)={\frac{2{N}_{c}}{{3\pi}^{2}}}{m}_{s}^{4}\lim
\limits_{\varepsilon\rightarrow 0}^{}\int_{1}^{\infty}
{\frac{{\left({{y}^{2}-1}\right)}^{3/2}}{1+{e}^{\varepsilon}
{e}^{{m}_{s}\beta y}}}dy
\label{eqA4}
\end{equation} 
where the infinitively small positive quantity
$\epsilon$ allows to regularize the
$T$ independent term of $P_{ideal\ gas} (s)$ (see below).  One then has
:
\begin{eqnarray} 
{P}_{ideal\
gas}\left({s}\right)&=&{\frac{{2N}_{c}}{{3\pi}^{2}}}{m}_{s}^{4}
\lim\limits_{\varepsilon\rightarrow 0}^{}\sum\limits_{n=1}^{\infty}
{\left({-1}\right)}^{n+1}{e}^{-n\varepsilon}\int_{1}^{\infty}
{\left({{y}^{2}-1}\right)}^{3/2}{e}^{-n{m}_{s}\beta y}dy\nonumber\\
&=&{\frac{{2N}_{c}}{{\pi}^{2}}}{m}_{s}^{4}\lim\limits_{\varepsilon
\rightarrow 0}^{}\sum\limits_{n=1}^{\infty} {\left({-1}\right)}^{n+1}
{e}^{-n\varepsilon}{\frac{{K}_{2}\left({n{m}_{s}\beta}\right)}
{{n}^{2}{m}_{s}^{2}{\beta}^{2}}}
\label{eqA5}
\end{eqnarray}
Writing the modified Bessel function $K_2 \ (z)$ as an ascending
series in power of $z$ \cite{AB72}
\begin{eqnarray}
{K}_{2}\left({z}\right)&=&2{z}^{-2}\left({1-{\frac{{z}^{2}}{4}}}
\right)\nonumber\\
&+&{\frac{1}{8}}{z}^{2}\sum\limits_{k=0}^{\infty}{\frac{\psi
\left({k+1}\right)+\psi\left({k+3}\right)}{k!\left({2+k}\right)!}}
{\left({{\frac{{z}^{2}}{4}}}\right)}^{k}-\ln{\frac{z}{2}}
{I}_{2}\left({z}\right),
\label{eqA6}
\end{eqnarray} 
with $\psi$ being the Digamma function and with the
other modified Bessel function $I_{2}(z)$
\begin{equation}
{I}_{2}\left({z}\right)={\frac{{z}^{2}}{4}}\sum\limits_{k=0}^{\infty}
{\frac{1}{k!\left({k+2}\right)!}}
{\left({{\frac{{z}^{2}}{4}}}\right)}^{k},
\label{eqA7}
\end{equation} 
one may write~(\ref{eqA5}) as a double sum which is
absolutely convergent, owing to the convergence factor
$e^{-n\epsilon}$.  One can easily control the first terms of the
expansion
\begin{eqnarray} 
{P}_{ideal\ gas}\left({s}\right)={p}_{4}{T}^{4}&+&{p}_{2}{T}^{2}+
{p}_{\ell}\ln\ T\nonumber\\ 
&+&{p}_{0}+{p}_{\ell -2}{T}^{-2}\ln\
T+{p}_{-2}{T}^{-2}+\ ...
\label{eqA8}
\end{eqnarray} 
One has
\begin{equation}
{p}_{4}={\frac{{4N}_{c}}{{\pi}^{2}}}\lim\limits_{\varepsilon
\rightarrow 0}^{}\sum\limits_{n=1}^{\infty}{\frac{{ e}^{-n\varepsilon}
{\left({-1}\right)}^{n+1}}{{n}^{4}}}={\frac{{4N}_{c}}{{\pi}^{2}}}\sum
\limits_{n=1}^{\infty}
{\frac{{\left({-1}\right)}^{n+1}}{{n}^{4}}}=
{\frac{7}{180}}{N}_{c}{\pi}^{2}
\label{eqA9}
\end{equation}
\begin{equation}
{p}_{2}={\frac{{N}_{c}}{{\pi}^{2}}}{m}_{s}^{2}\lim\limits_{\varepsilon
\rightarrow 0}^{}\sum\limits_{n=1}^{\infty}{\frac{{e}^{-n\varepsilon}
{\left({-1}\right)}^{n}}{{n}^{2}}}
=-{\frac{{N}_{c}}{{\pi}^{2}}}{m}_{s}^{2}\sum\limits_{n=1}^{\infty}
{\frac{{\left({-1}\right)}^{n+1}}{{n}^{2}}}
=-{\frac{{N}_{c}}{12}}{m}_{s}^{2}
\label{eqA10}
\end{equation}
\begin{eqnarray} 
{p}_{\ell}\ln\ T&=&{\frac{{N}_{c}}{{4\pi}^{2}}}{m}_{s}^{4}
\lim\limits_{\varepsilon\rightarrow 0}^{}\sum\limits_{n=1}^{\infty}
{\left({-}\right)}^{n}{e}^{-n\varepsilon}\ln{\frac{n{m}_{s}\beta}
{2}}\nonumber\\ 
&=&{\frac{{N}_{c}}{{4\pi}^{2}}}{m}_{s}^{4}\left[{\ln
{\frac{{m}_{s}\beta}{2}}\lim\limits_{\varepsilon\rightarrow
0}^{}\sum\limits_{n=1}^{\infty}{\left({-}
\right)}^{n}{e}^{-n\varepsilon}+
\lim\limits_{\varepsilon\rightarrow0}^{}\sum
\limits_{n=1}^{\infty}{\left({-}
\right)}^{n}{e}^{-n\varepsilon}\ln\ n}\right]\nonumber\\
&=&{\frac{{N}_{c}}{{4\pi}^{2}}}{m}_{s}^{4}\left[{-{\frac{1}{2}}\ln
{\frac{{m}_{s}\beta}{2}}+\lim\limits_{\varepsilon\rightarrow
0}^{}{\frac{d}
{dx}}\sum\limits_{n=1}^{\infty}
{\left({-}\right)}^{n}{e}^{-n\varepsilon}
{\left.{{n}^{x}}\right|}_{x=0}}\right]\nonumber\\
&=&{\frac{{N}_{c}}{{4\pi}^{2}}}{m}_{s}^{4}\left[{-{\frac{1}{2}}\ln
{\frac{{m}_{s}\beta}{2}}-{\frac{d}{dx}}{\left[{\left({{1-2}^{1+x}}
\right)\zeta\left({-x}\right)}\right]}_{x=0}}\right]\nonumber\\
&=&{\frac{{N}_{c}}{{4\pi}^{2}}}{m}_{s}^{4}\left[{-{\frac{1}{2}}
\ln{\frac{{m}_{s}\beta}{2}}+{\frac{1}{2}}\ln{\frac{\pi}{2}}}\right]
\nonumber\\
&=&-{\frac{{N}_{c}}{{8\pi}^{2}}}{m}_{s}^{4}
\ln{\frac{{m}_{s}\beta}{\pi}},
\label{eqA11}
\end{eqnarray}
where $\zeta$ is the Riemann $\zeta$ function,
\begin{eqnarray} 
{p}_{0}&=&{\frac{{N}_{c}}{{8\pi}^{2}}}{m}_{s}^{4}\lim
\limits_{\varepsilon\rightarrow 0}^{}{e}^{\varepsilon}\sum
\limits_{n=0}^{\infty}{\left({-1}\right)}^{n}
\left\{{\psi\left({1}\right)+\psi\left({3}\right)}\right\}
{e}^{-n\varepsilon}\nonumber\\
&=&{\frac{{N}_{c}}{{8\pi}^{2}}}{m}_{s}^{4}\left({-2\gamma+
{\frac{3}{2}}}\right)\lim\limits_{\varepsilon\rightarrow 0}^{}
{\frac{{1}}{1+{e}^{-\varepsilon}}}=
{\frac{{N}_{c}}{{16\pi}^{2}}}{m}_{s}^{4}\left({-2\gamma+
{\frac{3}{2}}}\right),
\label{eqA12}
\end{eqnarray}
where $\gamma$ is the Euler constant,
\begin{eqnarray} 
{p}_{\ell-2}{T}^{-2}\ln\ T&=&{\frac{{N}_{c}}{{48\pi}^{2}}}{m}_{s}^{6}
{T}^{-2}\lim\limits_{\varepsilon\rightarrow 0}^{}\sum
\limits_{n=1}^{\infty}{\left({-}\right)}^{n}{n}^{2}{e}^{-n\varepsilon}
\ln{\frac{{nm}_{s}\beta}{2}}\nonumber\\
&=&{\frac{{N}_{c}}{{48\pi}^{2}}}{m}_{s}^{6}{T}^{-2}\nonumber\\ 
& &\times\left[{\ln{\frac{{m}_{s}\beta}{2}}\lim\limits_{\varepsilon
\rightarrow 0}^{}\sum\limits_{n=1}^{\infty}{\left({-}\right)}^{n}
{e}^{-n\varepsilon}{n}^{2}+\lim\limits_{\varepsilon\rightarrow 0}^{}
\sum\limits_{n=1}^{\infty}{\left({-}\right)}^{n}{e}^{-n\varepsilon}
{n}^{2}\ln\ n}\right]\nonumber\\
&=&{\frac{{N}_{c}}{{48\pi}^{2}}}{m}_{s}^{6}{T}^{-2}
\left[{7\ln{\frac{{m}_{s}\beta}{2}}\zeta\left({-2}\right)-
{\frac{d}{dx}}{\left[{\left({{1-2}^{3+x}}\right)\zeta
\left({-2-x}\right)}\right]}_{x=0}}\right]\nonumber\\
&=&-{\frac{{7N}_{c}}{{48\pi}^{2}}}{m}_{s}^{6}{T}^{-2}\zeta'
\left({-2}\right),
\label{eqA13}
\end{eqnarray}
as the Riemann function $\zeta(s)$ vanishes for $s=-2$.  The $p_{-2}$
term vanishes, since it involves the sum
\begin{equation}
\lim\limits_{\varepsilon\rightarrow 0}^{}\sum\limits_{n=1}^{\infty}
{\left({-}\right)}^{n}{e}^{-n\varepsilon}{n}^{2}=\zeta\left({-2}
\right)=0.
\label{eqA14}
\end{equation} 
With~(\ref{eqA3}),~(\ref{eqA8})--(\ref{eqA14}), the high
temperature behaviour of the ideal gas pressure writes
\begin{eqnarray} 
P_{ideal\
gas}&=&{\frac{7}{60}}{N}_{c}{\pi}^{2}{T}^{4}-{\frac{{N}_{c}}{12}}
{m}_{s}^{2}{T}^{2}-{\frac{{N}_{c}}{{8\pi}^{2}}}{m}_{s}^{4}
\ln{\frac{{m}_{s}}{\pi
T}}+{\frac{{N}_{c}}{{16\pi}^{2}}}\left({{\frac{3}{2}}-2\gamma}\right)\
{m}_{s}^{4}\nonumber\\ 
& &-{\frac{{7N}_{c}}{{48\pi}^{2}}}{m}_{s}^{6}{\frac{\zeta'
\left({-2}\right)}{{T}^{2}}}+\ ...
\label{eqA15}
\end{eqnarray}
\begin{enumerate}
\item[b)] Energy density\\
\end{enumerate} 
Using the same technical procedure as before, one finds
:
\begin{equation}
\varepsilon_{ideal\ gas}={\varepsilon}_{ideal\ gas}\left({u}\right)+
{\varepsilon}_{ideal\ gas}\left({s}\right)
\label{eqA16}
\end{equation} 
with
\begin{equation} 
{\varepsilon}_{ideal\
gas}\left({u}\right)={\frac{{4N}_{c}}{{\pi}^{2}}}
\int_{0}^{\infty}{\frac{{k}^{3}}{1+{e}^{\beta k}}}dk=3{P}_{ideal\ gas}
\left({u}\right)={\frac{7}{30}}{N}_{c}{\pi}^{2}{T}^{4}
\label{eqA17}
\end{equation}
\begin{eqnarray} 
{\varepsilon}_{ideal\
gas}\left({s}\right)&=&{\frac{{2N}_{c}}{{\pi}^{2}}}
\int_{0}^{\infty}{k}^{2}{E}_{s}{\frac{1}{1+{e}^{\beta{E}_{s}}}}dk
\nonumber\\
&=&{\frac{{2N}_{c}}{{\pi}^{2}}}{m}_{s}^{4}\lim\limits_{\varepsilon
\rightarrow 0}^{}\int_{1}^{\infty}{\frac{\left[{{\left({{y}^{2}-1}
\right)}^{3/2}+{\left({{y}^{2}-1}\right)}^{1/2}}\right]}{1+{e}^{
\varepsilon}{e}^{{m}_{s}\beta y}}}dy\nonumber\\
&=&{\frac{{2N}_{c}}{{\pi}^{2}}}{m}_{s}^{4}\lim\limits_{\varepsilon
\rightarrow
0}^{}\sum\limits_{n=1}^{\infty}{e}^{-n\varepsilon}{\left({-1}
\right)}^{n+1}\left[{{\frac{{3K}_{2}\left({n{m}_{s}\beta}\right)}
{{n}^{2}{m}_{s}^{2}{\beta}^{2}}}+{\frac{{K}_{1}\left({{nm}_{s}\beta}
\right)}{n{m}_{s}\beta}}}\right]\nonumber\\ 
&=&3P_{ideal\
gas}+{\frac{2N_c}{\pi^2}{m}_{s}^{4}\lim\limits_{\varepsilon
\rightarrow 0}^{}\sum\limits_{n=1}^{\infty}{e}^{-n\varepsilon}
{\left({-}\right)}^{n+1}{\frac{{K}_{1}\left({n{m}_{s}\beta}
\right)}{n{m}_{s}\beta}}}.
\label{eqA18}
\end{eqnarray}
With \cite{AB72}
\begin{equation}
{K}_{1}\left({z}\right)={\frac{1}{z}}-{\frac{z}{4}}\sum
\limits_{k=0}^{\infty}
{\frac{\psi\left({k+1}\right)+\psi\left({k+2}\right)}{k!
\left({k+1}\right)!}}
{\left({{\frac{{z}^{2}}{4}}}\right)}^{k}+\ln{\frac{z}{2}}
{I}_{1}\left({z}\right)
\label{eqA19}
\end{equation} 
and
\begin{equation}
{I}_{1}\left({z}\right)={\frac{z}{2}}\sum\limits_{k=0}^{\infty}
{\frac{1}{k!\left({k+1}\right)!}}{\left({{\frac{{z}^{2}}{4}}}
\right)}^{k}
\label{eqA20}
\end{equation} 
one finds :
\begin{eqnarray} 
{\varepsilon}_{ideal\ gas}\left({s}\right)&=&{\varepsilon}_{4}{T}^{4}+
{\varepsilon}_{2}{T}^{2}+{\varepsilon}_{\ell}\ln\ T\ +{\varepsilon
}_{0}\nonumber\\ 
&+&{\varepsilon}_{\ell-2}{T}^{-2}\ln\
T+{\varepsilon}_{-2}{T}^{-2}+...
\label{eqA21}
\end{eqnarray}
with
\begin{equation}
{\varepsilon}_{4}={3p}_{4}={\frac{7}{60}}{N}_{c}{\pi}^{2}
\label{eqA22}
\end{equation}
\begin{equation}
{\varepsilon}_{2}=3p_2+{\frac{{N}_{c}}{{\pi}^{2}}}{m}_{s}^{2}\lim
\limits_{\varepsilon\rightarrow 0}^{}\sum\limits_{n=1}^{\infty}
{\frac{{e}^{-n\varepsilon}{\left({-1}\right)}^{n+1}}{{n}^{2}}}=-
{\frac{{N}_{c}}{12}}{m}_{s}^{2}={p}_{2}
\label{eqA23}
\end{equation}
\begin{eqnarray} 
{\varepsilon}_{\ell}\ln\ T&=&{3p}_{\ell}\ln\
T+{\frac{{N}_{c}}{{\pi}^{2}}}
{m}_{s}^{4}\lim\limits_{\varepsilon\rightarrow
0}^{}\sum\limits_{n=1}^{\infty}
{e}^{-n\varepsilon}{\left({-}\right)}^{n+1}\ln{\frac{{nm}_{s}\beta
}{2}}\nonumber\\
&=&{\frac{{N}_{c}}{{8\pi}^{2}}}{m}_{s}^{4}\ln{\frac{{m}_{s}\beta
}{\pi}}=-{p}_{\ell}\ln\ T,
\label{eqA24}
\end{eqnarray}
where we have used~(\ref{eqA11}),
\begin{eqnarray}
{\varepsilon}_{0}&=&{3p}_{0}-{\frac{{2N}_{c}}{{4\pi}^{2}}}
{m}_{s}^{4}\left({\psi\left({1}\right)+\psi\left({2}\right)}\right)
\lim\limits_{\varepsilon\rightarrow 0}^{}\sum\limits_{n=1}^{\infty}
{\left({-}\right)}^{n+1}{e}^{-n\varepsilon}\nonumber\\
&=&{\frac{{N}_{c}}{{16\pi}^{2}}}{m}_{s}^{4}\left({2\gamma+
{\frac{1}{2}}}\right),
\label{eqA25}
\end{eqnarray}
\begin{eqnarray} 
{\varepsilon}_{\ell-2}{T}^{-2}\ln\
T&=&{3p}_{\ell-2}{T}^{-2}\ln\ T\nonumber\\
&+&{\frac{{2N}_{c}}{{8\pi}^{2}}}{m}_{s}^{6}{T}^{-\ 2}\lim
\limits_{\varepsilon\rightarrow 0}^{}\sum\limits_{n=1}^{\infty}
{\left({-}\right)}^{n+1}{e}^{-n\varepsilon}{n}^{2}
\ln{\frac{{nm}_{s}\beta}{2}}\nonumber\\ 
&=&3{p}_{\ell-2}{T}^{-2}\ln\
T+{\frac{7}{8}}{\frac{{N}_{c}}{{\pi}^{2}}}
{m}_{s}^{6}{T}^{-2}\zeta'\left({-2}\right)\nonumber\\
&=&{\frac{{7N}_{c}}{{16\pi}^{2}}}{m}_{s}^{6}{T}^{-2}\zeta'
\left({-2}\right)=-3{p}_{\ell-2}{T}^{-2}\ln\ T.
\label{eqA26}
\end{eqnarray}
The term $\varepsilon_{-2}$ vanishes because it is proportional to
$\zeta
\left(-2\right)$.  Gathering the results, one obtains :
\begin{eqnarray} 
{\varepsilon}_{ideal\
gas}={\frac{7}{20}}{N}_{c}{\pi}^{2}{T}^{4}&-&
{\frac{{N}_{c}}{12}}{m}_{s}^{2}{T}^{2}+{\frac{{N}_{c}}{{8\pi}^{2}}}
{m}_{s}^{4}\ln{\frac{{m}_{s}}{\pi
T}}+{\frac{{N}_{c}}{{8\pi}^{2}}}{m}_{s}^{4}
\left({\gamma+{\frac{1}{4}}}\right)\nonumber\\
&+&{\frac{7}{16}}{\frac{{N}_{c}}{{\pi}^{2}}}{m}_{s}^{6}\zeta'
\left({-2}\right){T}^{-2}+...
\label{eqA27}
\end{eqnarray}
\begin{enumerate}
\item[c)] Entropy density\\
\end{enumerate} 
Using~(\ref{eq22}),~(\ref{eqA15}) and~(\ref{eqA27}),
one gets :
\begin{equation}
Ts=2{\frac{{N}_{c}}{{\pi}^{2}}}{m}_{s}^{4}\lim\limits_{\varepsilon
\rightarrow 0}^{}\sum\limits_{n=1}^{\infty}{\left({-1}\right)}^{n+1}
{e}^{-n\varepsilon}{\frac{{K}_{3}\left({n{m}_{s}\beta}\right)}
{n{m}_{s}\beta}}
\label{eqA28}
\end{equation}
\begin{equation}
s={\frac{7}{15}}{N}_{c}{\pi}^{2}{T}^{3}-{\frac{{N}_{c}}{6}}{m}_{s}^{2}
T+{\frac{{N}_{c}}{{8\pi}^{2}}}{m}_{s}^{4}{T}^{-1}+\frac{7}{24}\frac{N_c}
{\pi^2}m_{s}^{6}T^{-3}\zeta'
\left(-2\right)+...
\label{eqA29}
\end{equation}
\end{appendix}

\pagebreak

\pagebreak

\section*{\bf Figure captions}

{\bf Figure 1.}\par 
Chiral phase transition $\mu_{sc}$ versus $\mu_{uc}$ for
$T=0$ MeV (a), $T=100$ MeV (b), $T=161$ MeV (c).  The free parameters
of the model are ${M}_{u}^{0}=600$ MeV, ${\chi}_{0}=125$ MeV.  The line
with the crosses (open circles) corresponds to a vanishing $\bar{s}s$
($\bar{u}u$) condensate.

{\bf Figure 2.}\par 
Temperature behaviour of the constituent $u$- and $s$-quark masses and
of the gluon condensate for
${\mu}_{u}={\mu}_{s}=250$ MeV (a) and ${\mu}_{u}=200$ MeV,
${\mu}_{s}=0$ MeV (b) with
${M}_{u}^{0}=600$ MeV, ${\chi}_{0}=125$ MeV.

{\bf Figure 3.}\par 
Chiral first order phase transition (${T}_{c}$, ${\mu}_{uc}$,
${\mu}_{sc}$) : artistic view.

{\bf Figure 4.}\par 
Temperature behaviour of the constituent $u$- and $s$-quark masses and
of the gluon condensate for
${\mu}_{u}={\mu}_{s}=0$ MeV (a) and ${\mu}_{u}={\mu}_{s}=250$ MeV (b)
with ${M}_{u}^{0}=300$ MeV, 
${\chi}_{0}=80$ MeV.

{\bf Figure 5.}\par 
Chiral phase transition $\mu_{sc}$ versus $\mu_{uc}$ for
$T=0$ MeV (a), $T=100$ MeV (b), $T=130$ MeV (c) and $T=165$ MeV (d),
with
${M}_{u}^{0}=300$ MeV, ${\chi}_{0}=80$ MeV.  Same conventions as for
Fig.~1.

{\bf Figure 6.}\par 
Chiral phase transition
$\left({{T}_{c},{\mu}_{uc}, {\mu}_{sc}}\right)$, summarizing the results
presented in Fig.~5 : artistic view.

{\bf Figure 7.}\par 
Equation of state for ${\mu}_{u}={\mu}_{s}=0$ MeV (a) and
${\mu}_{u}={\mu}_{s}=250$ MeV (b).  Approximated $T^4$ linear fits
above $T_c$ are indicated in (a) while quadratic fits in $T^2$ are
required for part (b).  Results are shown for
${M}_{u}^{0}=600$ MeV, ${\chi}_{0}=125$ MeV (crosses) and
${M}_{u}^{0}=300$ MeV, ${\chi}_{0}=80$ MeV (full dots).

{\bf Figure 8.}\par 
Energy density versus $T^4$ for ${\mu}_{u}={\mu}_{s}=0$ MeV (a) and
${\mu}_{u}={\mu}_{s}=250$ MeV (b).  Same conventions as in Fig.~7.

{\bf Figure 9.}\par 
Entropy density versus $T^3$ for ${\mu}_{u}={\mu}_{s} =0$ MeV (a) and
${\mu}_{u}={\mu}_{s}=250$ MeV (b).  Approximated $T^3$ linear fits,
above $T_c$ are indicated in (a) while quadratic fits for $Ts$ are
required for part (b).  Same conventions as in Fig.~7.

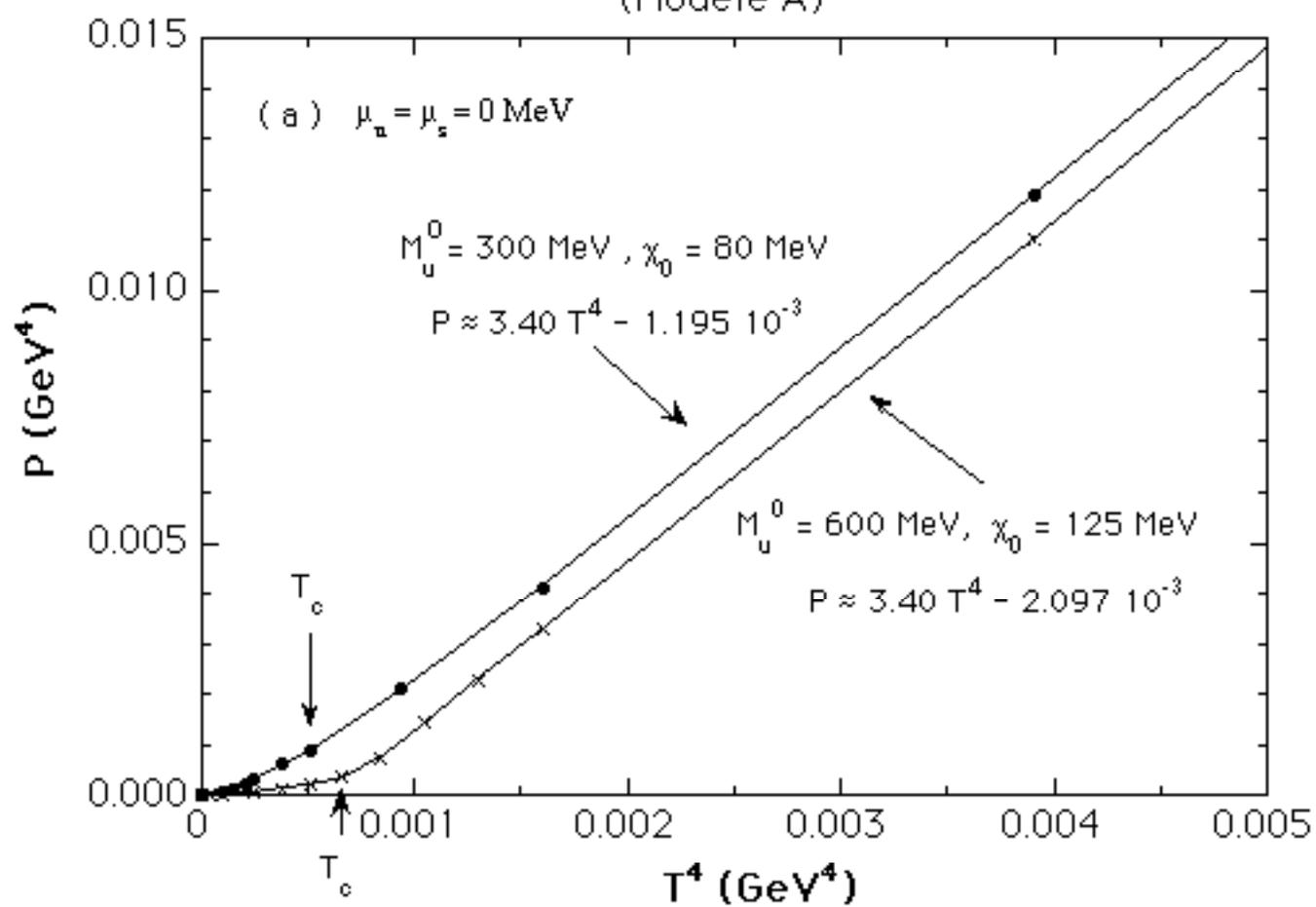
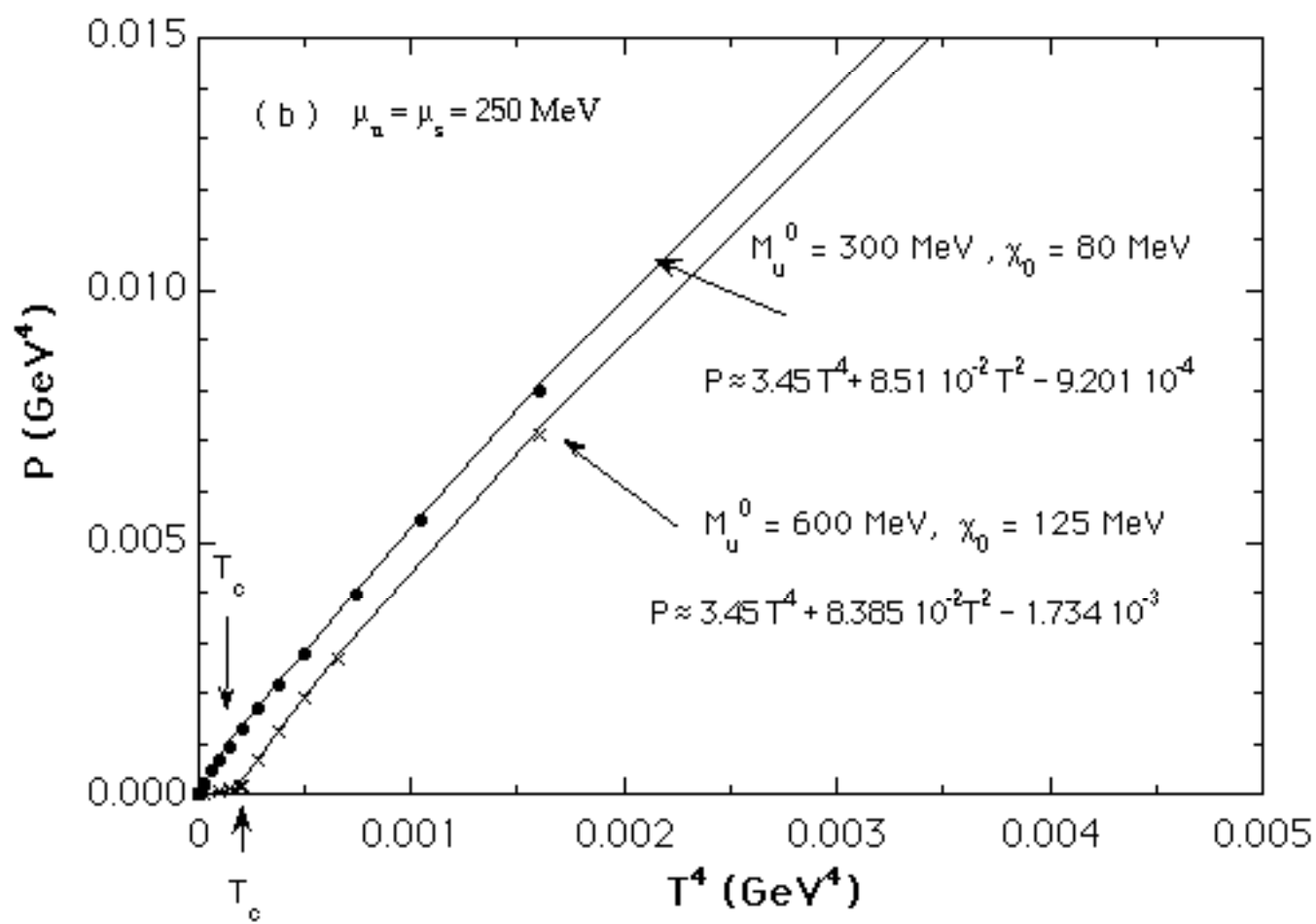
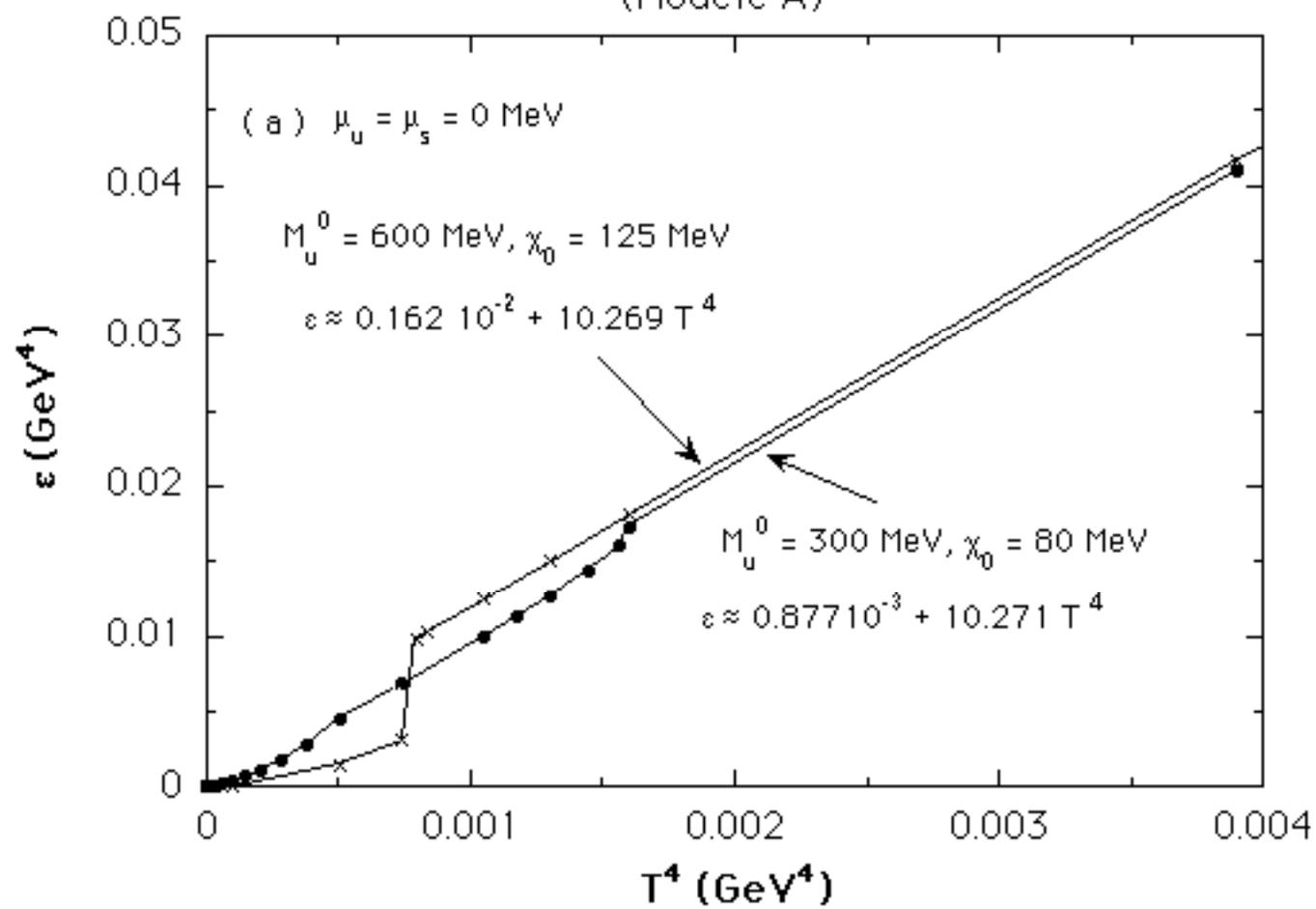
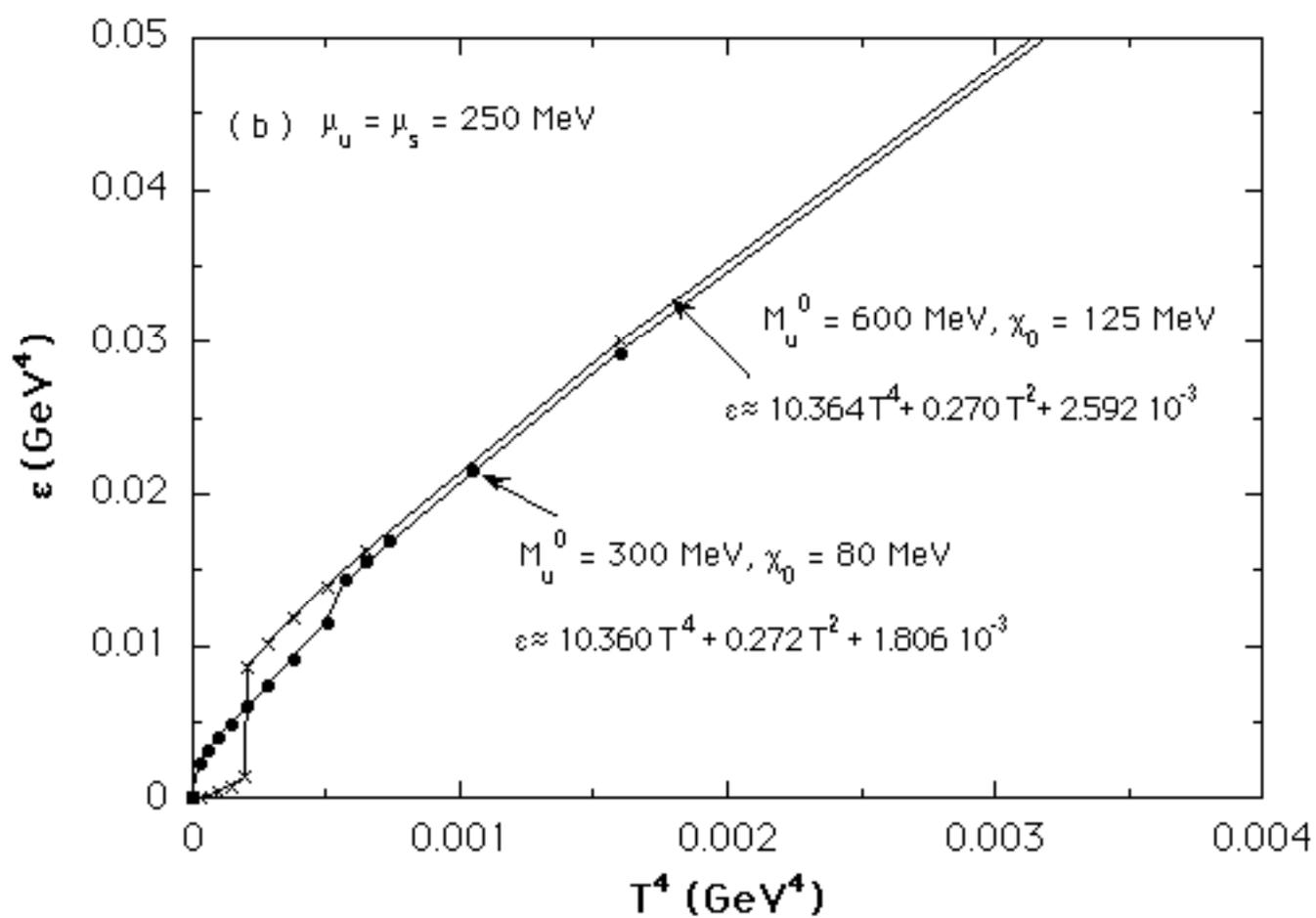
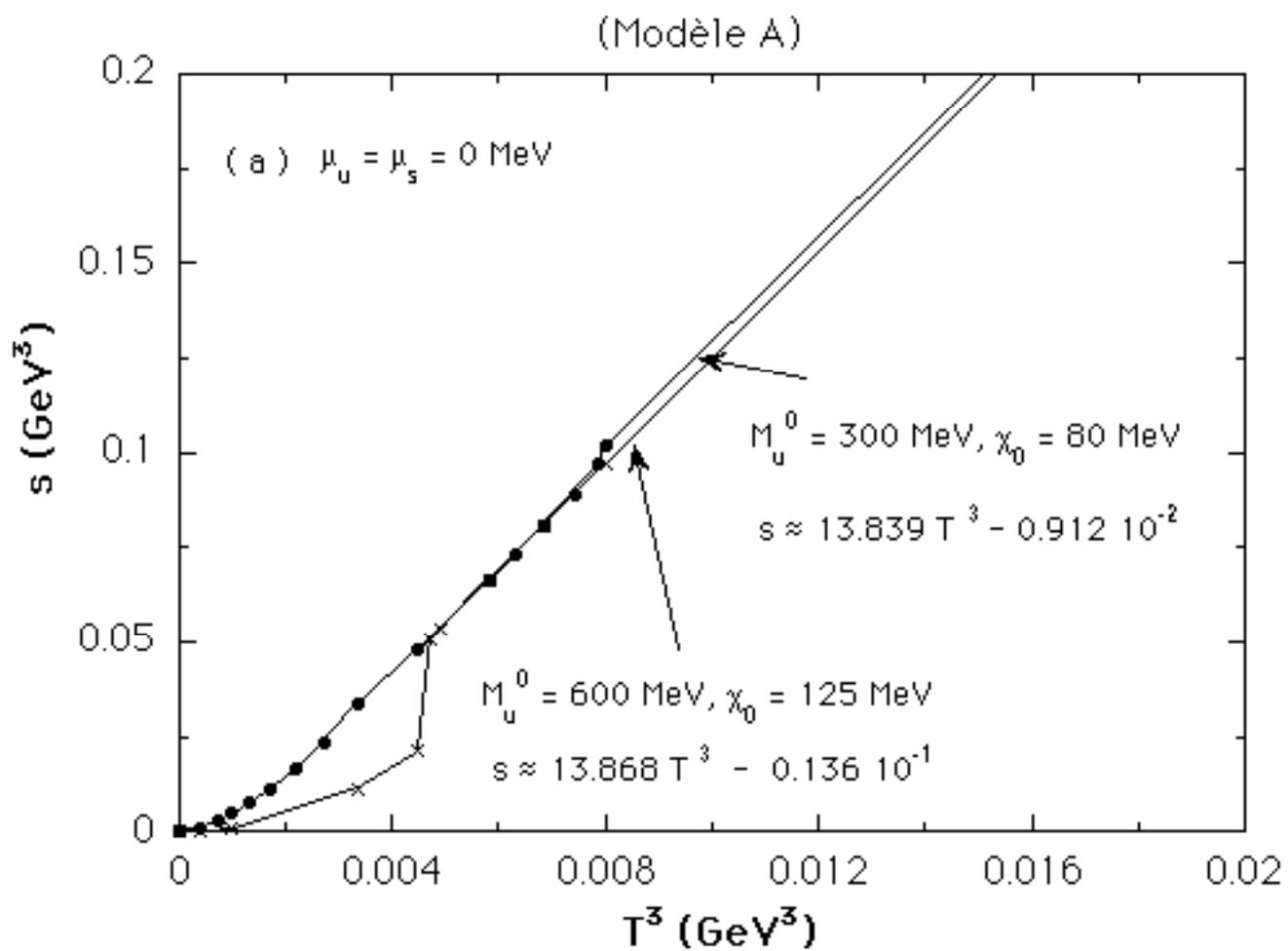
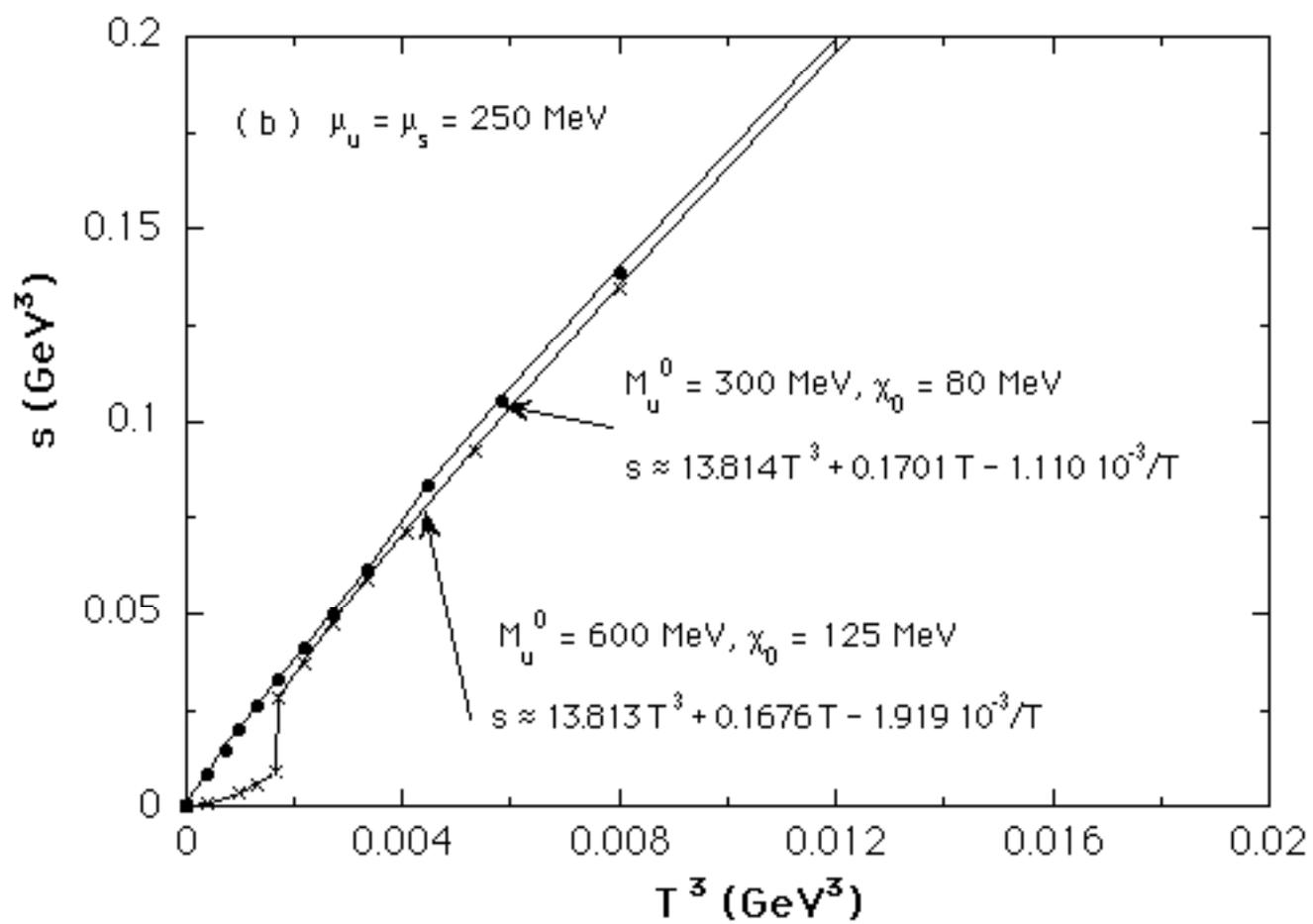
\end{document}